\documentclass{PoS}

\usepackage{latexsym}
\usepackage{graphicx}
\usepackage{epstopdf}
\usepackage{bm}
\title{Nucleon structure functions from dynamical (2+1)-flavor domain wall fermions}

\ShortTitle{2+1f DWF nucleon structure function}

\author{\speaker{Shigemi Ohta} (RBC and UKQCD Collaborations)\\
     Inst.\ Particle and Nuclear Studies, KEK, Tsukuba, Ibaraki 305-0801, Japan\\
	Physics Department, SOKENDAI, Hayama, Kanagawa 240-0193, Japan\\
	RIKEN BNL Research Center, BNL, Upton, NY 11973, USA\\
	E-mail: \email{shigemi.ohta@kek.jp}}
        
\abstract{
We report lattice-volume independence of low moments of nucleon structure functions from the coarse RIKEN-BNL-Columbia (RBC) and UKQCD joint dynamical (2+1)-flavor domain-wall fermions (DWF) ensembles at the lattice cut off of \(a^{-1}\sim1.7\) GeV.
The isovector quark momentum fraction, \(\langle x \rangle_{u-d}\), and helicity fraction, \(\langle x \rangle_{\Delta u - \Delta d}\), both fully non-perturbatively renormalized are studied on two spatial volumes of \((\sim {\rm 2.7 fm})^3\) and \((\sim {\rm 1.8 fm})^3\).
Their naturally renormalized ratio, \(\langle x \rangle_{u-d}/\langle x \rangle_{\Delta u - \Delta d}\), is not affected by any finite-size effect.
It does not depend strongly on light quark mass and does agree well with the experiment.
The respective absolute values, fully non-perturbatively renormalized, do not show any finite-size effect either.
They show trending toward the respective experimental values at the lightest up- and down-quark mass.
This trending down to the experimental values appears to be a real physical effect driven by lighter quarks.
The observations are in contrast to the huge finite-size effect seen in the axial-current form factors.

\vspace{-176mm}\parbox{\textwidth}{\flushright\large\rm \hfill KEK-TH-1335, RBRC-813}\vspace{176mm}
}

\FullConference{The XXVII International Symposium on Lattice Field Theory - LAT2009\\
		 July 26-31 2009\\
		 Peking University, Beijing, China}

\begin{document}

\section{Introduction}

We report lattice-volume independence of low moments of nucleon structure functions.  
In Lattice 2008 we reported our lattice numerical calculations of some low moments of nucleon structure functions \cite{Ohta:2008kd} with the RBC and UKQCD joint dynamical (2+1)-flavor domain-wall fermions (DWF) ensembles at the lattice cut off of \(a^{-1}\sim1.7\) GeV \cite{Allton:2008pn}.
In contrast to companion form factor calculations \cite{Yamazaki:2008py,Yamazaki:2009zq}, we did not see any sign of the huge finite-size effect in the structure functions.
However those calculations then were complete with only the larger, \(({\rm 2.7 fm})^3\) volume.
Now we have completed the calculation with the smaller, \(({\rm 1.8 fm})^3\) volume, and are more confident that the structure functions do not suffer from the finite-size effect so much as the form factors.

The nucleon structure functions are measured in deep inelastic scattering of leptons off nucleons \cite{Bloom:1969kc,Breidenbach:1969kd}, the cross section of which is factorized in terms of leptonic and hadronic tensors,
\(
\propto
l^{\mu\nu}W_{\mu\nu}.
\)
Since the leptonic tensor, \(l_{\mu\nu}\), is known, the cross section provides us with structure information about the target nucleon through the hadronic tensor, $W_{\mu\nu}$, which is decomposed into symmetric unpolarized and antisymmetric polarized parts:
\begin{equation}\label{eq:Wsym}
W^{\{\mu\nu\}}(x,Q^2) =
\left( -g^{\mu\nu} + \frac{q^\mu
q^\nu}{q^2}\right)  {F_1(x,Q^2)} +
\left(P^\mu-\frac{\nu}{q^2}q^\mu\right)\left(P^\nu-\frac{\nu}{q^2}q^\nu\right)
\frac{F_2(x,Q^2)}{\nu},
\end{equation}
\begin{equation}\label{eq:Want}
W^{[\mu\nu]}(x,Q^2) = i\epsilon^{\mu\nu\rho\sigma} q_\rho
\left(\frac{S_\sigma}{\nu}({g_1(x,Q^2)} \right. +
 \left. {g_2(x,Q^2)}) - \frac{q\cdot S
P_\sigma}{\nu^2}{g_2(x,Q^2)} \right),
\end{equation}
with kinematic variables defined as $\nu = q\cdot P$, $S^2 = -M^2$, and $x=Q^2/2\nu$, and \(Q^2=|q^2|\).
The unpolarized structure functions are $F_1(x,Q^2)$ and $F_2(x,Q^2)$, and the polarized, $g_1(x,Q^2)$ and $g_2(x,Q^2)$.
Their moments are described in terms of Wilson's operator product expansion:
\begin{eqnarray}\label{eq:moments}
2 \int_0^1 dx\,x^{n-1} {F_1(x,Q^2)} &=& \sum_{q=u,d}
c^{(q)}_{1,n}\: \langle x^n \rangle_{q}(\mu)
+{O(1/Q^2)},
\nonumber \\
\int_0^1 dx\,x^{n-2} {F_2(x,Q^2)} &=& \sum_{f=u,d}
c^{(q)}_{2,n}\: \langle x^n \rangle_{q}(\mu)
+{O(1/Q^2)},
\nonumber \\
2\int_0^1 dx\,x^n {g_1(x,Q^2)}
  &=& \sum_{q=u,d} e^{(q)}_{1,n}\: \langle x^n \rangle_{\Delta q}(\mu)
+{O(1/Q^2)},  \nonumber
 \\
2\int_0^1 dx\,x^n {g_2(x,Q^2)}
  &=& \frac{1}{2}\frac{n}{n+1} \sum_{q=u,d} \left[e^{q}_{2,n}\: d_n^{q}(\mu)
-  2 e^{q}_{1,n}\: \langle x^n \rangle_{\Delta
q}(\mu)\right] + {O(1/Q^2)},
\end{eqnarray}
where we suppressed the \((\mu^2/Q^2,g(\mu))\) dependence of the perturbatively known Wilson coefficients, $c_1$, $c_2$, $e_1$, and $e_2$.
The moments, ${\langle x^n \rangle_{q}(\mu)}$, ${\langle x^n \rangle_{\Delta q}(\mu)}$ and $d_n(\mu)$ are calculable on the lattice as forward nucleon matrix elements of certain local operators.


In the following we report our lattice numerical calculations of two lowest isovector moments of the structure functions, the quark momentum fraction, \(\langle x\rangle_{u-d}(\mu)\), the helicity fraction, \(\langle x\rangle_{\Delta u-\Delta d}(\mu)\), and their ratio.
We note the results for the tensor charge, \(\langle 1\rangle_{\delta u - \delta d}(\mu)\), which probes transverse spin structure of nucleon \cite{Anselmino:2007fs,RHICSpin}, and twist-3 coefficient \(d_1\) of the \(g_2\) polarized structure function, were reported in Lattice 2008 \cite{Ohta:2008kd}.

\section{Formulation}

We use the standard proton operator, \(B = \epsilon_{abc} (u_a^T C \gamma_5d_b)u_c\) to create and annihilate proton states.
We Gaussian-smear this operator for better overlap with the ground state with both zero and finite momentum, as we conducted a companion form-factor calculations \cite{Yamazaki:2008py,Yamazaki:2009zq}.
Since the up and down quark mass are degenerate in these calculations, the isospin symmetry is exact.
This is of course a well-known good approximation.
We project the positive-parity ground state, so our two-point proton function takes the form
\begin{equation}
C_{\rm 2pt}(t) =
\sum_{\alpha,\beta}
\left(\frac{1+\gamma_t}{2}\right)_{\alpha\beta} 
\langle B_\beta(t_{\rm sink})\overline{B}_\alpha(t_{\rm source}) \rangle,
\end{equation}
with \(\displaystyle  t=t_{\rm sink}-t_{\rm source}\).
We insert an appropriate observable operator \(O(\vec{q}, t')\) at time \(t'=t_{\rm source}+\tau\), \(0 \le \tau \le t\), and possibly finite momentum transfer \(\vec{q}\), to obtain a form factor or structure function moment three-point function,
\begin{equation}
C^{\Gamma, O}_{\rm 3pt}(t, \tau, \vec{q}) =
\sum_{\alpha,\beta}
\Gamma_{\alpha\beta}
\langle B_\beta(t_{\rm sink}) O(\vec{q}, t') \overline{B}_\alpha(t_{\rm source}) \rangle,
\end{equation}
with appropriate projection, \(\displaystyle \Gamma=\frac{1+\gamma_t}{2}\), for the spin-unpolarized, and \(\displaystyle \Gamma=\frac{1+\gamma_t}{2}i\gamma_5\gamma_k, k\ne 4\), for the polarized.
Ratios of these two- and three-point functions give plateaux for \(0<\tau<t\) that give the bare lattice matrix elements of desired observables: e.g.\ 
\(\displaystyle
\langle O\rangle^{\rm bare}=
\frac{C^{\Gamma, O}_{\rm 3pt}(t, \tau) }{C_{\rm 2pt}(t)}
\)
at \(q^2=0\).
We renormalize the structure function moments by Rome-Southampton RI-MOM non-perturbative renomalization prescription \cite{Dawson:1997ic}.
The good continuum-like flavor and chiral symmetries of domain-wall fermions \cite{Kaplan:1992bt,Shamir:1993zy,Furman:1994ky} are very useful here in eliminating unwanted lattice-artifact mixings that are present in many other fermion schemes.
See our earlier publications, ref.\ \cite{Lin:2008uz}, and references cited there in for further details.

The RBC-UKQCD joint (2+1)-flavor dynamical DWF coarse ensembles \cite{Allton:2008pn} are used for the calculations.
These ensembles are generated with Iwasaki action at the coupling \(\beta=2.13\) which corresponds to
the lattice cut off of about \(a^{-1}=1.73(3)\) GeV.
There are two lattice volumes, \(16^3\times 32\) and \(24^3\times 64\), corresponding to linear spatial extent of  about 1.8 and 2.7 fm.
The dynamical strange and up and down quarks are described by DWF actions with the fifth-dimensional mass of \(M_5=1.8\).
The strange mass is set at 0.04 in lattice unit and turned out to be about twelve percent heavier than physical including the additive correction of the residual mass, \(m_{\rm res}=0.003\).
The degenerate up and down mass is varied at 0.03, 0.02, 0.01 and 0.005. 
We summarize the gauge configurations and pion and nucleon mass of each ensemble in Table \ref{table:statistics}.
\begin{table}[t]
\begin{center}
\begin{tabular}{lrrcll} \hline
\multicolumn{1}{c}{$m_fa$} & 
\multicolumn{1}{c}{\# of config.'s} &
\multicolumn{1}{c}{meas.\ interval} &
\multicolumn{1}{c}{$N_{\rm sources}$} &
\multicolumn{1}{c}{$m_\pi$ (GeV)} &
\multicolumn{1}{c}{$m_N$ (GeV)}\\ \hline
0.005 & 646+286 & 10 & 4 & 0.3294(13) &1.154(7)\\ 
0.01   & 356 & 10 & 4 & 0.4164(12) &1.216(7)\\
0.02   &  98  & 20 & 4 & 0.5550(12) &1.381(12)\\ 
0.03   & 106 & 20 & 4 & 0.6681(15) &1.546(12)\\ \hline
\end{tabular}
\end{center}
\caption{Number of gauge configurations and pion and nucleon mass.\label{table:statistics}}
\end{table}

There are two important sources of systematic error: finite spatial size of the lattice and excited states contamination:
Chiral-perturbation-inspired analysis of the former for meson observables suggests the dimensionless product, \(m_\pi L\), of the calculated pion mass, \(m_\pi\) and lattice linear spatial extent, \(L\), should be set greater than 4 to drive the finite-volume correction below one percent, and the available lattice calculations seem to support this.
While our present parameters satisfy this condition, it should be emphasized that this criterion is not known sufficient for baryon observables.
Indeed our companion calculations on nucleon isovector form factors \cite{Yamazaki:2008py,Yamazaki:2009zq} strongly suggests much larger volumes with \(m_\pi L \sim 6-7\) are necessary.
Here we test this by performing structure function calculations on two lattice volumes, \((\sim {\rm 2.7 fm})^3\) and \((\sim {\rm 1.8 fm})^3\)

One should adjust the time separation, \(t\), between the nucleon source and sink appropriately so the resultant nucleon observables are free of contamination from excited states.
The separation has to be made longer as we set the quark masses lighter.
Here our previous study with two dynamical flavors of DWF quarks \cite{Lin:2008uz} help as the lattice cutoff is similar at 1.7 GeV:
By comparing the isovector quark momentum fraction, \(\langle x\rangle_{u-d}\), at the up/down mass of 0.02, where pion mass is about 500 MeV and nucleon 1.3 GeV, a clear systematic difference is seen between the shorter time separation of 10, or about 1.16 fm, and longer 12, or 1.39 fm: the former averages to 0.236(9) while the latter 0.195(17).
Although the former appears statistically better, as the two-point function that provides the normalization has not decayed much yet, it is not manifestly free of excited-state contamination that is expected more at shorter time separation.
The latter suffers statistically, as the two-point function that provides the normalization has decayed, but is certainly freer of excited-state contamination than the former: Of the two, we have to choose the latter.
If larger statistics is required, then we have to collect as much statistics as necessary for the latter.
Ideally we should collect more statistics at even larger source-sink separation to confirm the excited-state contamination has been eliminated by the separation of 12.

Since in the present ensembles we explore much lighter up/down quark mass than this two-flavor example, with pion mass as light as 330 MeV and nucleon 1.15 GeV, the source/sink time separation must be chosen longer: longer than 1.2 fm at least.
For the lightest up/down quark mass, \(m_fa=0.005\), in the present (2+1)-flavor dynamical DWF calculation, the nucleon signal begins to decay at \(t=12\), or about 1.4 fm \cite{Blum:2008kd} (see Figure \ref{fig:Nemass}):
\begin{figure}[b]
\begin{center}
\includegraphics[width=\textwidth,clip]{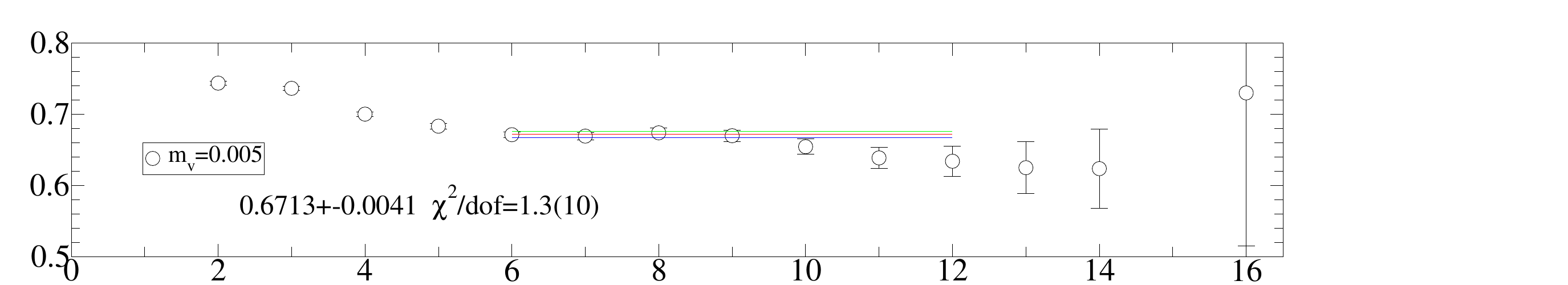}
\end{center}
\caption{Nucleon effective mass at quark mass \(m_fa=0.005\) in the present calculation with (2+1)-flavor dynamical DWF quarks.  The nucleon signal begins to decay at time separation 12.  By which time we are as free of excited state as possible while maintaining the nucleon ground state.\label{fig:Nemass}}
\end{figure}
hence it is about the longest distance we can choose without losing the signal, and hence about as free of excited-state contamination as we can go.
We found the bare three-point function signals for quark momentum and helicity fractions for this source-sink separation of \(t=12\) are acceptable \cite{Ohta:2008kd}.
Thus we choose the source-sink separation of 12, or about 1.4 fm for the present work, as the two-flavor example suggests strongly that anything shorter would suffer from excited-state contamination.
To confirm this choice is sufficiently long to eliminate the excited-state contamination remains a future problem.  

\section{Results}

The ratio, \(\langle x \rangle_{u-d}/\langle x \rangle_{\Delta u - \Delta d}\), of the isovector quark momentum fraction to the helicity fraction, is naturally renormalized in DWF lattice QCD because the two quantities share a common renormalization \cite{Orginos:2005uy}:
the momentum fraction, \(\langle x \rangle_{u-d}\), the first moment of the \(F_{1,2}\) unpolarized structure functions, and helicity fraction, \(\langle x \rangle_{\Delta u - \Delta d}\), the first moment of the \(g_1\) polarized structure function, are related by a chiral rotation and the DWF preserve the chiral symmetry.
Thus the ratio calculated on the lattice is directly comparable with the experiment.
\begin{figure}
\begin{center}
\includegraphics[width=.48\textwidth,clip]{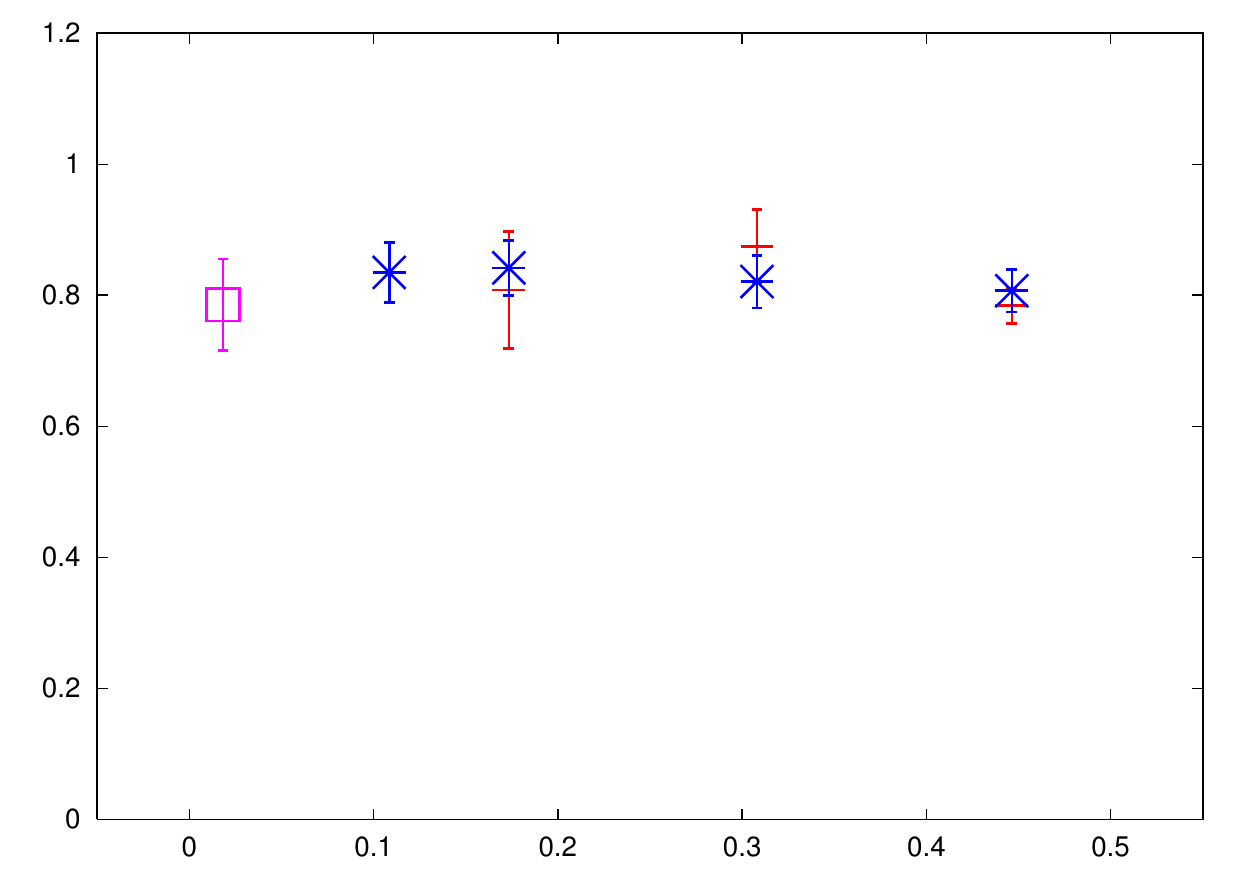}
\end{center}
\caption{
Ratio, \(\langle x\rangle_{u-d}/\langle x\rangle_{\Delta u-\Delta d}\), of isovector momentum and helicity fractions, naturally renormalized on the lattice, from \(({\rm 2.7 fm})^3\) (\(\times\)) \(({\rm 1.8 fm})^3\) (\(+\)), plotted against pion mass squared in \({\rm GeV}^2\).
They are in excellent agreement with the experiment which is denoted by a square (\(\Box\)).
No discernible dependence on volume nor pion mass can be detected.
\label{fig:xqxDq}}
\end{figure}
Our results for this ratio do not show any discernible dependence on the up/down quark mass, and are in excellent agreement with the experiment (see Figure \ref{fig:xqxDq}.) 
This is in contrast to the similarly naturally renormalized ratio of \(g_A/g_V\) of elastic form factors which at the lightest point deviates significantly from heavier mass results and the experiment \cite{Yamazaki:2008py}.
This suggests individual moments of inelastic structure functions such as the momentum fraction, \(\langle x \rangle_{u-d}\), and helicity fraction, \(\langle x \rangle_{\Delta u - \Delta d}\), may not suffer so severely from the finite-size effect that plagues elastic form factor calculations.

\begin{figure}
\begin{center}
\includegraphics[width=.48\textwidth,clip]{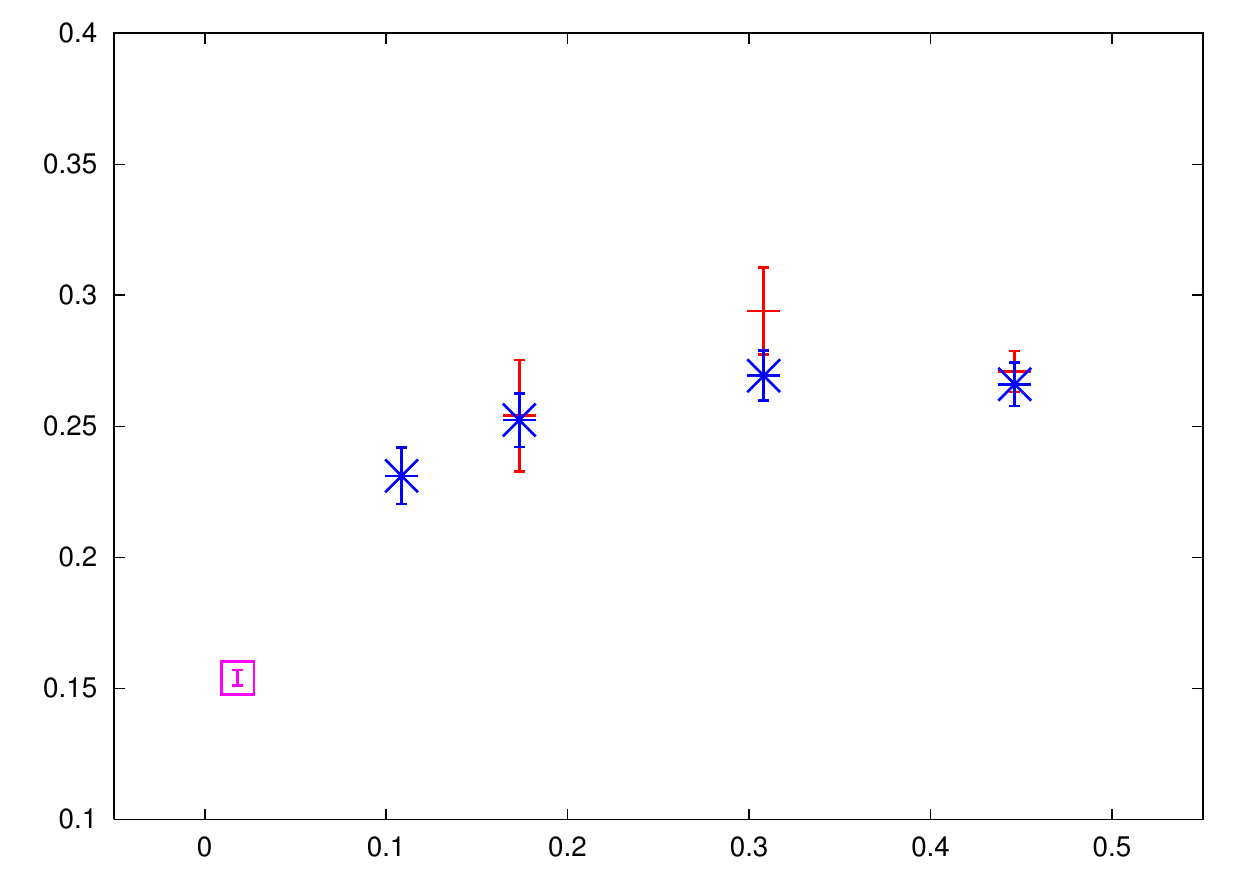}
\includegraphics[width=.48\textwidth,clip]{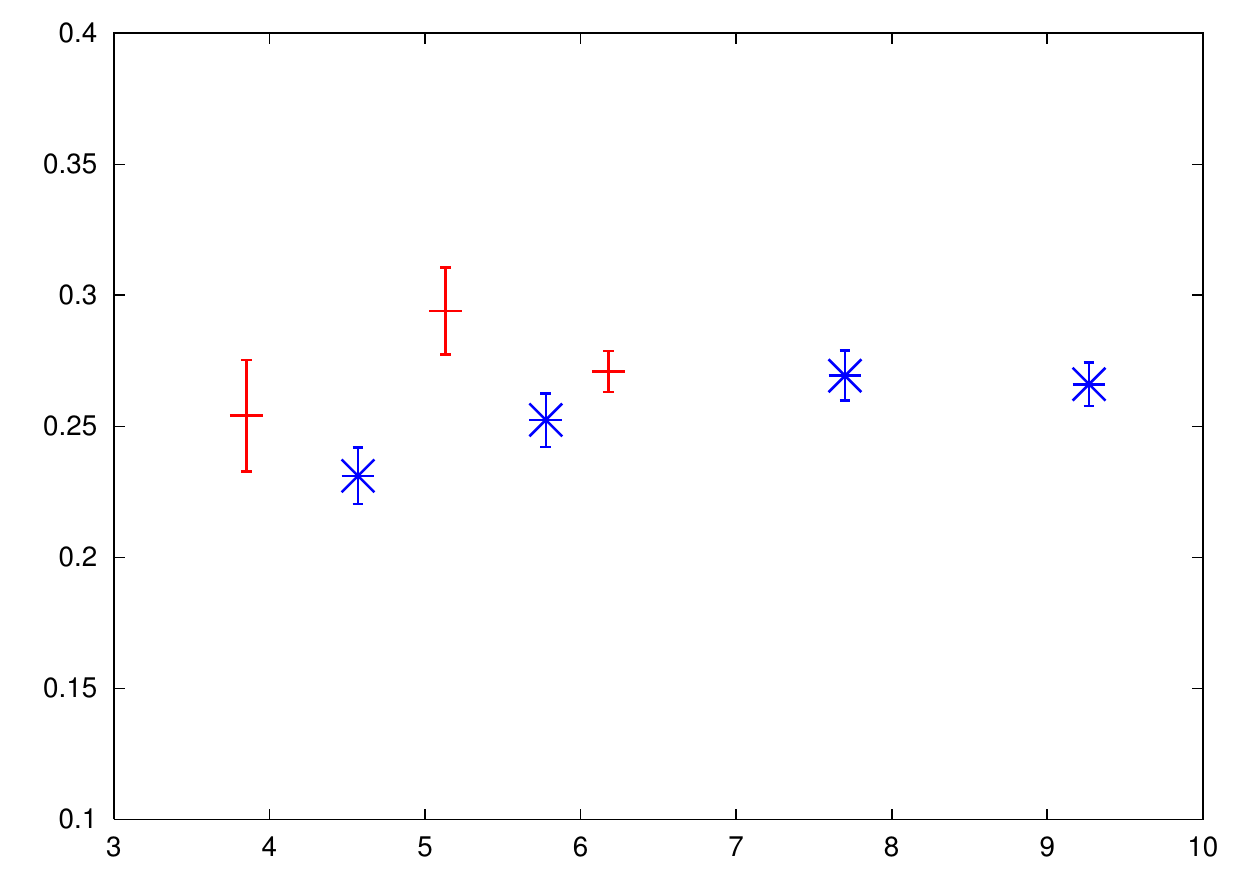}
\end{center}
\caption{Absolute value of isovector quark momentum fraction, \(\langle x \rangle_{u-d}\), fully non-perturbatively renormalized.
Plotted against the pion mass squared in \({\rm GeV}^2\) (left) in comparison with the experiment which is denoted by a square (\(\Box\)), and against \(m_\pi L\) (right).
Results from the larger, \(L=2.7\) fm volume are described by \(\times\) symbols and the smaller \(L=1.8\) fm volume by \(+\).
\label{fig:xq}}
\end{figure}

The absolute value of the isovector quark momentum fraction, \(\langle x \rangle_{u-d}\), fully non-perturbatively renormalized and run to \(\overline{\rm MS}\) at 2 GeV, \(Z^{\overline{\rm MS}}({\rm 2 GeV}) = 1.15(4)\), is shown in Figure \ref{fig:xq}.
On the larger, 2.7 fm, volume, the three heavier points stay roughly the constant which is about 70 \% higher at \(\sim 0.26\) than the experiment, about 0.15.
This behavior is not so different from old RBC quenched results \cite{Orginos:2005uy} wtih similar up/down quark mass.
The lightest point shows a sign of deviation away from this heavy constant behavior, but in contrast to the form factor deviations, is trending toward the experiment.
Since lighter quark can more easily share its momentum with other degrees of freedom, this trending toward the experiment may well be a real physical effect:
It does not have to be affected by the finite spatial size of the lattice.
This question is now clarified by the calculations on the small, \(({\rm 1.8 fm})^3\), volume:
at  \(m_fa=0.01\) on the smaller volume of 1.8 fm, with \(m_\pi L\) smaller than that at the \(m_fa=0.005\) on 2.7 fm volume, the calculated momentum fraction is in agreement with the large volume result at the same mass.
This absence of the finite-size effect scaling in \(m_\pi L\) can also be seen in the right pane of the figure where the same quantities are plotted against this scaling variable.

\begin{figure}[b]
\begin{center}
\includegraphics[width=.48\textwidth,clip]{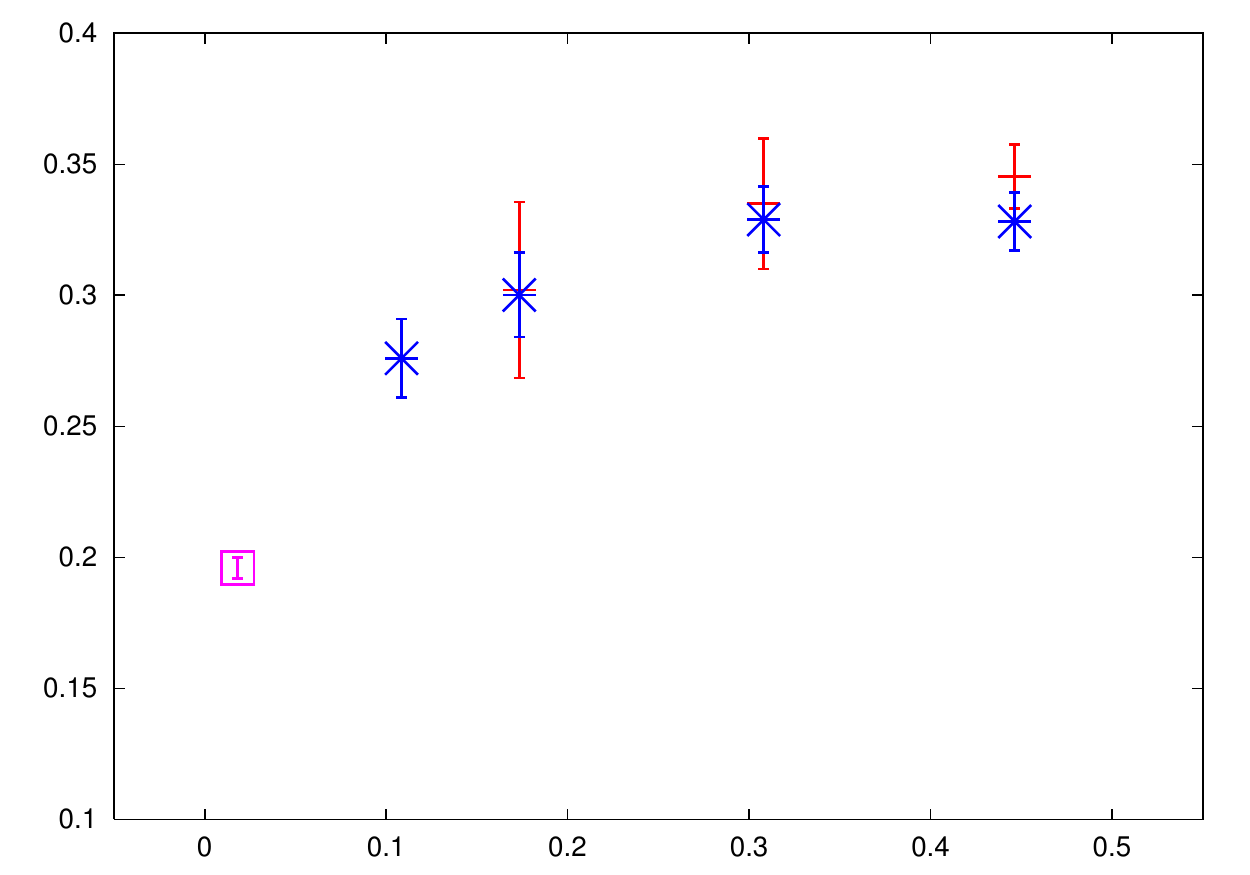}
\end{center}
\caption{Absolute value of isovector quark helicity fraction, \(\langle x \rangle_{\Delta u-\Delta d}\), fully non-perturbatively renormalized.
Plotted against the pion mass squared in \({\rm GeV}^2\)  in comparison with the experiment which is denoted by a square (\(\Box\)). 
Results from the larger, \(L=2.7\) fm volume are described by \(\times\) symbols and the smaller \(L=1.8\) fm volume by \(+\).
\label{fig:xDq}}
\end{figure}

The isovector quark helicity fraction, \(\langle x \rangle_{\Delta u - \Delta d}\), is also fully non-perturbatively renormalized and run to \(\overline{\rm MS}\) at 2 GeV (See Figure \ref{fig:xDq}.)
The renormalization, \(Z^{\overline{\rm MS}}({\rm 2 GeV}) = 1.15(3)\), agrees very well with the momentum fraction one.
As can be expected from the constant behavior of the ratio, \(\langle x \rangle_{u-d}/\langle x \rangle_{\Delta u - \Delta d}\), this denominator observable exhibits very similar behavior to the numerator one discussed in the above: the three heavier points stay roughly the constant which is about 70 \% higher than the experiment.
The lightest point then shows a sign of deviation away from this heavy constant behavior and trends toward the experiment.
This trending toward the experiment is independent of the lattice volume and of the form factor scaling variable, \(m_\pi L\):
It is likely a real physical effect arising from light quark mass.

\section{Conclusions}

A naturally renormalized observable with DWF quarks because of well-preserved chiral symmetry, the ratio, \(\langle x \rangle_{u-d}/\langle x \rangle_{\Delta u - \Delta d}\), of the isovector quark momentum and helicity fractions, does not depend strongly on the degenerate up- and down-quark mass and does agree well with the experiment.
This is in contrast to a similarly naturally renormalized ratio, the axial charge, \(g_A/g_V\), where a huge finite-size effect takes the calculation away from the experiment at ligher ud-quark mass values \cite{Yamazaki:2008py} .
The absolute values of the fractions, \(\langle x \rangle_{u-d}\) and \(\langle x \rangle_{\Delta u - \Delta d}\), fully non-perturbatively renormalized, show trending toward the respective experimental values at the lightest ud-quark mass value, away from constancy at heavier values.
Comparison of the results on the two different lattice volumes 
does not reveal any finite-size effect.
In particular, we rule out huge finite-size effect scaling in \(m_\pi L\) seen in the axial charge \cite{Yamazaki:2008py}.
Thus the trending down to the respective experimental values of these two isovector fractions appears to be a real physical effect driven by lighter ud-quarks, in contrast to the finite-size effect in the axial-current form factors.

I thank J.~Zanotti, T.~Yamazaki, and other RBC and UKQCD members. 
RIKEN, BNL, US DOE, Edinburgh University and UK PPARC provided facilities essential for this work.

\end{document}